# Finding influential users of an online health community: a new metric based on sentiment influence[*]

Kang Zhao[1,2], Greta Greer[3], Baojun Qiu[2,4], Prasenjit Mitra[2], Kenneth Portier[3], and John Yen[2]

[1] University of Iowa, S224 PBB, Iowa City, IA 52242. kang-zhao@uiowa.edu.

[2] College of Information Sciences and Technology, The Pennsylvania State University, University Park, PA 16802. {pmitra, jyen}@ist.psu.edu.

[3] American Cancer Society, 250 Williams Street NW, Atlanta, GA, 30303. {greta.greer, kenneth.portier}@cancer.org.

[4] eBay Inc. 2065 Hamilton Ave, San Jose, CA 95125. baqiu@ebay.com.

## Abstract

What characterizes influential users in online health communities (OHCs)? We hypothesize that (1) the emotional support received by OHC members can be assessed from their sentiment expressed in online interactions, and (2) such assessments can help to identify influential OHC members. Through text mining and sentiment analysis of users' online interactions, we propose a novel metric that directly measures a user's ability to affect the sentiment of others. Using dataset from an OHC, we demonstrate that this metric is highly effective in identifying influential users. In addition, combining the metric with other traditional measures further improves the identification of influential users. This study can facilitate online community management and advance our understanding of social influence in OHCs.

[*] This research is supported by American Cancer Society and the Pennsylvania State University. Kang Zhao and Baojun Qiu conducted part of this research during their doctoral studies at the Pennsylvania State University.

# 1. Introduction

As more and more people use the Internet to satisfy their health-related needs, many of them seek support through participation in an online health community (OHC) where they interact with peers facing similar health problems. According to a study by the Pew Research Center, 80% of adult Internet users in the U.S. use Internet for health-related purposes. Among them, 34% reads about health-related experiences or comments from others (1) and 5% of all Internet users participated in an OHC (2). Obtaining psychosocial support is one of the key benefits of the participation in OHC (3, 4). The effectiveness and proper functioning of these communities may be affected by the presence and activities of influential users (IUs), who provide psychosocial supports to other members of the community and have "the power or capacity of causing an effect in indirect or intangible ways" (5). However, an influential user may disappear from an OHC due to his/her health condition (e.g., recurrent cancer), which can present major challenges for the OHC. Hence, the identification of current and emerging influential users can help to improve the sustainability of OHCs.

We propose a novel approach to IU identification based on the assumption that through their online activities influential OHC users are able to affect the emotion of other community members. Hence, we aim at identifying IUs in an OHC by (1) measuring the effect of inter-personal influence, (2) identifying key contributors to the influence in threaded discussions, and (3) aggregating a person's contribution to social influence in the community. The proposed approach utilizes individual OHC users' sentiment dynamics and develops a new metric based on sentiment influence. This approach is applied to data from the online forum of a peer-support community sponsored by the American Cancer Society, the Cancer Survivors Network (CSN

http://csn.cancer.org). The dataset used contains 48,779 threaded discussions with more than 468,000 posts from 27,173 de-identified users over a 10-year period ending in October, 2010. Each thread starts with an initial post, which is published by the thread originator and followed by responding replies from other users (respondents). In many cases, the thread will contain additional posts from the originator (self-replies).

## 2. The Proposed Approach

### 2.1 Sentiment Analysis.

In an OHC, user emotions cannot be directly observed, but the sentiment of their posts can reflect their emotions at the time of posting. Manually labeling sentiment for so many posts is not feasible. Instead, our previous research designed an algorithm to detect the sentiment of posts automatically and classify texts into *positive* or *negative* sentiment classes (6). To calibrate the classification algorithm, we manually label 298 randomly selected posts as belonging to *positive* or *negative* sentiment classes. Examples of initial posts and responding posts with negative and positive sentiment are shown in Table 1.

**Table 1.** Examples of posts in CSN and their sentiment classes.

| | | |
|---|---|---|
| Negative sentiment | Initial post | My mom became resistant to chemo after 7 treatments and now the trial drug is no longer working :(, ... |
| | Reply | I feel really sorry for that. I know how painful it is… |
| Positive sentiment | Initial post | Hooray! The tumor is gone according to my doctor! … |
| | Reply | …, I love the way you think, ..., hope is crucial and no one can deny that a cure may be right around the corner!!! |

Next, we extract several lexical and style features from the content of each post, including the number of words with positive sentiment (e.g., "happy" and "joy") and negative sentiment (e.g., "disappointed" and "painful"), the number of Internet slang (e.g., "LOL" and ":-)"), the numbers of question marks and exclamation marks, etc. These features are chosen because they can differentiate posts with positive sentiment and others with negative sentiment. For instance, a post that expresses negative sentiment often contains many words with negative sentiment. Finally, machine-learning-based classifiers are trained based on these manually labeled posts (with cross validations). The ultimate goal is having a classifier that is able to assign to each post the correct sentiment labeled by human experts.

Of the several classifiers we tried, AdaBoost (7) with regression trees as weak learners has the best sentiment classification performance (8). The classifier achieves an accuracy of 79.2%, meaning that the classifier can correctly determine sentiment labels for about 80% of the 298 manually labeled posts. This performance is in line with other sentiment analysis of various domains that have reported accuracy rates ranging from 66% to 84% (6, 9). Then, this sentiment classification model is applied to all unlabeled posts, producing a sentiment label for each. Specifically, for each post $p_i$, the sentiment classification model estimates a sentiment posterior probability, $P_r(c=pos/p_i)$, which measures how likely it is that the post belongs to the *positive* sentiment class given its post characteristics. If $P_r(c=pos/p_i)>0.5$, post $p_i$ is labeled as *positive*; otherwise, it is labeled as *negative*. Figure 1 illustrates the process of sentiment classification for posts. Materials and Methods S1 includes more detailed descriptions of our sentiment classification approaches.

**2.2 The new metric.**

Given the assigned sentiment of all the posts in the OHC, we utilize the sentiment dynamics within threads to develop a metric that reflects each user's ability to influence others' sentiments. Thread originators often start a thread to seek support from the community on a health-related issue. Replies from other users exert some level of influence on the originator's feeling on the issue, so that sentiment of the originator's subsequent self-replies in the thread may change. From such sentiment change, we derive a measure of how influential responding users are.

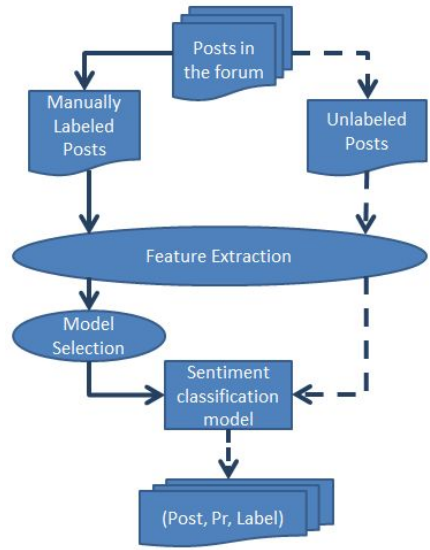

**Fig1.** The process of sentiment classification for posts in the OHC.

If a thread does not receive any responding reply, presumably the thread originator does not receive support from the community with respect to that thread. If the thread originator does not post any reply in a thread she started (i.e., a self-reply), we cannot measure her change of sentiment on the issue discussed in this thread. Among the 48,779 threads, only 23,000 threads have at least one responding reply, and contain at least one self-reply from the thread originator. By comparing a thread originator's sentiment in the initial post with her sentiment in subsequent self-replies, it is possible to measure the influence from thread respondents.

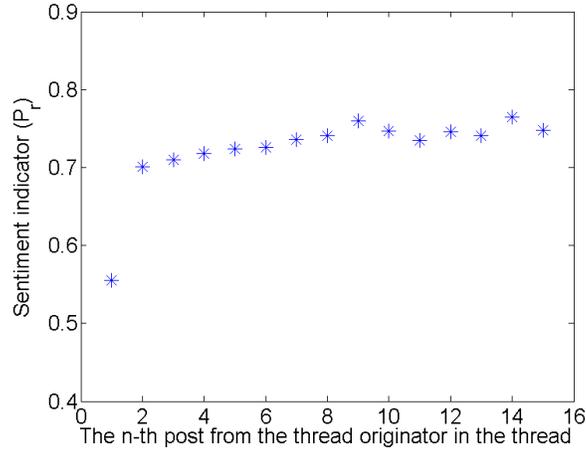

**Fig.2.** Sentiment change of thread originators by number of posts. A point represents the average sentiment of thread originators' n-th posts in threads they initiated. As the 2nd post from the originator is the 1st self-reply, the 2nd data point from the left-hand side denotes the average sentiment of originators' first self-replies.

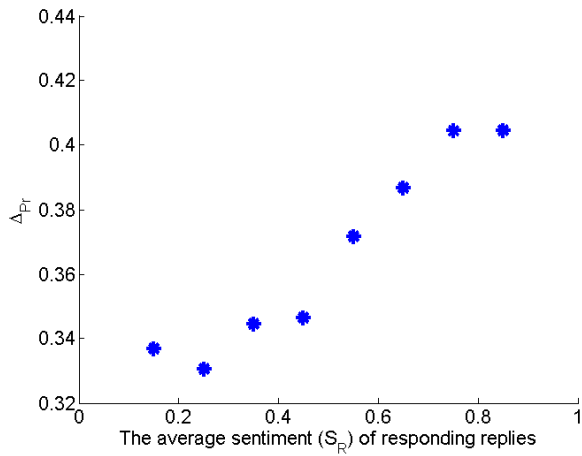

**Fig.3.** Change in originators' sentiment as a function of the average sentiment of responding replies.

As Figure 2 illustrates, sentiment of thread originators, on average, changes within threads they initiated. The most significant change in average sentiment occurs between their initial posts and their first self-replies. Their sentiment does not change much between their first self-reply and subsequent posts in the same thread. For our analysis, the sentiment of originators' self-replies is simply averaged as $S_F = \sum_{i=1}^{N} P_r(c = pos | s_i) / N$, where $s_i$ refers to one of the $N$ self-replies from the thread originator. Similarly, the sentiment of responding replies is averaged as $S_R=$

$\sum_{j=1}^{M} P_r(c = pos | r_j)/M$, where $r_j$ is one of the $M$ responding replies in the thread. Then the sentiment change indicator for a thread originator is computed as $\Delta_{Pr} = S_F - S_0$, where $S_0 = P_r(c=pos/p_0)$ is the sentiment from the initial post of the thread. Plotting $\Delta_{Pr}$ against $S_R$ (Figure 3) demonstrates that $\Delta_{Pr}$ tends to have higher values as $S_R$ increases. While the curve does demonstrate some non-linearity, the Pearson correlation coefficient between $\Delta_{Pr}$ and $S_R$ is 0.96 (P-value≤0.0001).

We conclude that the more positive the sentiment of responding replies, the greater the positive change in the originator's sentiment. While this finding only establishes association, it does lend support to our thesis that the sentiment of thread respondents can impact that of originators. The analysis also establishes that social support from respondents generally influences thread originators in a positive way. After at least one responding reply is received, about 75% of all the thread originators who started with negative sentiment expressed positive subsequent sentiment; among those who started with positive sentiment, 85% stayed positive in their subsequent sentiment. Figure 4 shows the distribution of $\Delta_{Pr}$ for originators who start a thread with a post that has negative sentiment. 7.9% of them have $\Delta_{Pr} < 0$. The average $\Delta_{Pr}$ is 0.1359 and a t-test concludes that $\Delta_{Pr}$ is significantly greater than 0.

Having demonstrated that the sentiment of respondents has an impact on the change in sentiment of the thread initiator, we return to the issue of identifying influential users. We posit that influential users post greater numbers of influential responses and use the number of influential responding replies (IRR) as a metric of influence. An IRR is a responding reply that is able to affect the sentiment of posts by the thread originator. While all responding replies in a thread may alter the sentiment of originators' self-replies, we only consider responding replies that are pub-lished before an originators' first self-reply within the thread. The rationale for this is twofold. First, a thread originator's sentiment often changes significantly between the initial post and the first self-reply, but it only changes little afterwards (See Figure 2). Hence, responses in a thread before the originator posts again are more likely to be influential than other responses. Second, the temporal intervals between initial posts and first self-replies have a median value of 17 hours, with two-thirds of them below 24 hours. The cumulative distributions of temporal intervals between initial posts and the first/last self-reply (Figure 5) shows that threads can sometimes last quite a long time with the originator often keeping the conversation going with multiple self-replies. Thus by focusing on those early responses, we can further increase the probability that an originator's sentiment change is actually due to the responding replies and not due to other events happening offline, such as changes in their physical conditions, personal support from friends and acquaintances, or approaching holidays.

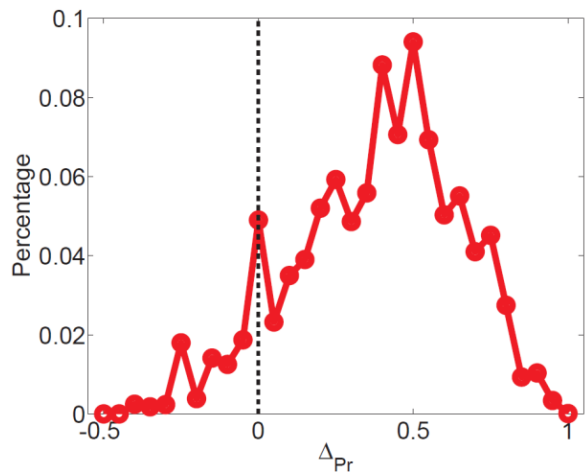

**Fig.4.** the distribution of $\Delta_{Pr}$ for originators who start a thread with a post that has negative sentiment

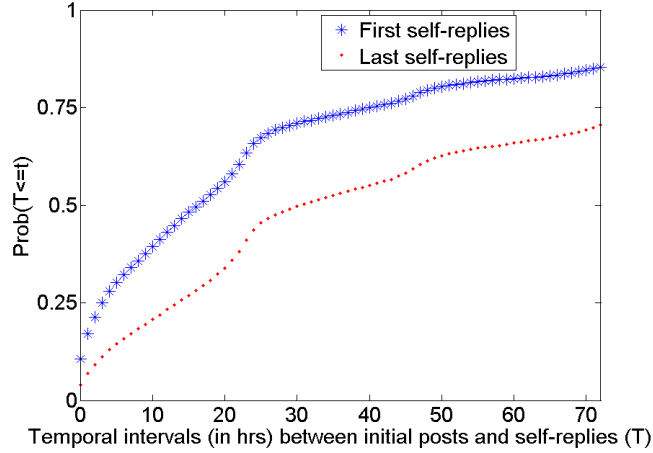

**Fig. 5.** Cumulative distribution of the intervals between initial posts and their first/last self-replies.

Operationally, an IRR moves the thread originator's sentiment in the direction of the reply's sentiment (positive or negative). If an IRR $r_j$ brings in positive sentiment to a thread (i.e., our sentiment classifier assigns $P_r(c=pos/r_j)>0.5$ for post $r_j$), the originator's sentiment in the first self-reply ($s_1$) should become more positive compared to that in the initial post ($p_0$), i.e., $P_r(c=pos/s_1)>P_r(c=pos/p_0)$; if $r_j$ contains negative sentiment (i.e., $P_r(c=pos/r_j)<0.5$), the originator's sentiment in the first self-reply should become more negative than it was in the initial post, i.e., $P_r(c=pos/s_1)<P_r(c=pos/p_0)$. If there are multiple IRRs between the initial post and the first self-reply, all are considered to be contributory to the originator's sentiment change. This is analogous to the aggregation of influence from multiple actors in the threshold model (10, 11), even though thresholds that represent individual differences on the ease of being influenced are not explicit in the definition of IRR. Formally, a responding reply $r_j$ in the thread started by initial post $p_0$ is an IRR if and only if the following two conditions are met:

> Condition 1: $T(p_0)<T(r_j)<T(s_1)$, where $T(p)$ is the publishing time of post $p$.
> Condition 2: $P_r(c=pos/r_j)>0.5$, if $P_r(c=pos/s_1)>P_r(c=pos/p_0)$
> or $P_r(c=pos/r_j)<0.5$, if $P_r(c=pos/s_1)<P_r(c=pos/p_0)$;

*Please note that the metric considers responding replies that bring both positive and negative sentiment to thread originators.* There are at least two reasons to consider negative influences on sentiments. First, in some contexts, it is not always appropriate to publish responding replies with positive sentiment. For example, if a thread reports the death of a community member, influential users may show their sympathy in the thread that may deepen the sadness of the originator (We examined threads where IRRs had negative sentiment impacts on thread originators. A preliminary lexical search found that 13% of the initial posts contain words or expressions related to death. Please refer to Material and Methods S3 for details). Responding replies like this can also be supportive in this special context and do not necessarily have a negative impact on the community. Second, a small number of the influential users may feel so passionate about an opinion that they, while contributing to the emotion support of some members, could annoy those who do not agree with their opinions. Distinguishing different types of negative sentiment influence and finding IUs whose activities may have negative impacts on the whole community is beyond the scope of this research.

In general, the number of IRRs posted by a user is a reflection of the individual's engagement in and contribution to the community, promptness in providing support, and more importantly, the level of influence that this user can exert on others. These are all important characteristics of influential users in the OHC.

## 3. Evaluation

One is able to rank users according to their numbers of IRRs--the higher the number of IRRs is, the more likely they are influential users. Validation of the ranking requires an independently derived and ranked list of OHC influential users. Unfortunately, while it is easy to label users whose activity levels are very low as non-influential users, identifying a true influential user requires good knowledge of each user's history of activities and of other users' reactions to their posts over an extended period of time. With the help of domain experts, in this case the CSN community manager and two staff members who monitor forum content on a full-time basis, 41 community members (referred to as IU List-1) are nominated as influential users. Ranking of List-1 is not performed because these experts do not think that it is feasible to do so reliably for such a high number of users[1].

Although List-1 comes from subjective evaluations and does not include all the IUs in the community, it provides a starting point for evaluating the utility of using IRR to identify IUs. Because we do not know all the IUs in the community, we cannot directly use traditional metrics, such as precision and recall, to measure the performance of our approach. Instead, we evaluate our ranking using Top-*K* recalls (also known as Recall@K): we check how many of the 41 nominated influential users in List-1 can be ranked within top *K* (with various *K* values) by our new IRR metric. If *n* of them are ranked within top *K*, then the Top-*K* recall is *n/41*. The higher the Top-*K* recall for a ranking metric, the better the performance of the metric. Table 2 also lists the Top-*K* recalls of various user rankings by traditional metrics, including contribution metrics and network centralities (Here we use a post-reply network, where there is an edge pointing from user *A* to user *B* if *A* published a responding reply in a thread started by *B*). Intuitively, higher *K* values lead to higher Top-*K* recalls. More importantly, the performance of our new IRR metric is better than that of all other metrics for top-50 recall, top-100 recall, and top-150 recall.

---

[1] Note that in information retrieval and search engine evaluation, only relevance judgments are sought and not totally ordered ranks because of the same reason.

**Table 2.** Compare the Top-K recall from various single-metric user rankings (using IU List-1).

| Metric | K=50 | K=100 | K=150 |
|---|---|---|---|
| Total number of threads initiated | 0.342 | 0.439 | 0.585 |
| Total number of posts | 0.415 | 0.707 | 0.781 |
| In-degree in the post-reply network | 0.317 | 0.512 | 0.610 |
| Out-degree in the post-reply network | 0.390 | 0.659 | 0.780 |
| Betweenness in the post-reply network | 0.293 | 0.366 | 0.488 |
| PageRank in the post-reply network | 0.390 | 0.561 | 0.732 |
| Early replies within 24 hours | 0.487 | 0.707 | 0.781 |
| Total number of IRRs | 0.512 | 0.732 | 0.805 |

**Table 3.** Compare the performance of the IRR ranking and an ensemble classifier (using both IU List-1 and List-2)[2].

| | K=50 | | K=100 | | K=150 | |
|---|---|---|---|---|---|---|
| | Recall (max =0.397) | Precision | Recall (max =0.794) | Precision | Recall | Precision (max=0.840) |
| IRR Ranking | 0.349 | 0.880 | 0.627 | 0.790 | 0.762 | 0.640 |
| The original ensemble classifier | 0.278 | 0.700 | 0.532 | 0.670 | 0.698 | 0.587 |
| The new ensemble classifier with IRR | 0.373 | 0.940 | 0.579 | 0.730 | 1.000 | 0.840 |

How does the performance of the IRR metric compare to the combined power of traditional metrics? Our early work developed several classifiers (12) that can be used to also identify IUs. These classifiers utilized 68 user features that measured users' contributions in various ways (e.g., the numbers of posts and active days), network centralities (e.g., degree, betweenness, and PageRank), and post content (e.g., the frequency of words with positive/negative sentiment in a user's posts). The best performing classifier utilized an ensemble random-forest model built on outcomes from other individual classifiers fitted to the same online community. Materials and Methods S2 includes more information about the ensemble classifier. To assess the precision of this ensemble classifier, domain experts reviewed a new list of users that were ranked within the top 150 by the classifier but were not included in List-1. Using criteria similar to those that gen-erated IU List-1, domain experts endorsed an additional 85 influential users (referred to as IU List-2). The longer list of IUs endorsed by domain experts enables better comparison between our IRR ranking and the classifier.

---

[2] Note that there are 85+41=126 influential users when we combine List-1 and List-2. As 126>100>50, the maximum possible values for Top-50 and Top-100 recalls are not 1. For instance, even though the top 50 users are all influential ones, the maximum possible Top-50 recall is only 50/126=0.397. Similarly, the maximum Top-100 recall is 100/126=0.794; the maximum Top-150 precision is 126/150=0.84.

Surprisingly, as Table 3 shows, the performance of our IRR ranking is better than that of the classifier in both Top-*K* recalls and precisions, even though the IRR ranking only uses one metric and the other classifier uses 60 features. In addition to recall, the high precision of our metric is very useful. For example, the Top-50 precision is 0.880, meaning that 88% of the top 50 users are indeed influential ones.

To test the validity of the IRR metric, we also conduct sensitivity analysis. The sentiment class label of a post is the basis for finding IRRs in the OHC. Originally, a classification threshold of $P_r(c=pos/p_i)=0.5$ is used to determine the sentiment of posts, with those above 0.5 being classified as having positive sentiment. Then is the new IRR metric robust when we pick different threshold values for sentiment classification? We vary the classification threshold values from 0.3 to 0.7 and count each user's number of IRRs again. It turns out that the IRR metric is very robust against such threshold changes. Users' numbers of IRRs based on different threshold values are still highly correlated with correlation coefficients higher than 0.998 (see Table 4). In other words, despite of the changing threshold values, a user with many IRRs still is a high-IRR user, and vice versa. Consequently, the IRR ranking maintains consistent performance in identifying influential users when different threshold values are used (see Table 5).

To further improve the identification of influential users and illustrate the synergistic benefit of our new metric, we also incorporate the IRR metric as a new feature into the original ensemble classifier. The IRR-enhanced classifier performs much better than the previous one (See the last row in Table 3). Its strong performance in Top-50 recall and precision are especially desirable for finding members with very high influence in the OHC. The new classifier's perfect Top-150 recall means that it could find all of the nominated and endorsed influential users within top 150. The imperfect top-150 precision is also acceptable because the combined list with 126 IUs (from IU List-1 and List-2) still may not include all influential users in the OHC.

**Table 4.** Correlation coefficients for the numbers of IRRs with different thresholds values for sentiment classification.

| Threshold values | Correlation coefficient (p-value) |
|---|---|
| $P_r(c=pos/p_i)=0.5$ vs $P_r(c=pos/p_i)=0.3$ | 0.9985 (0.0000) |
| $P_r(c=pos/p_i)=0.5$ vs $P_r(c=pos/p_i)=0.4$ | 0.9986 (0.0000) |
| $P_r(c=pos/p_i)=0.5$ vs $P_r(c=pos/p_i)=0.6$ | 0.9995 (0.0000) |
| $P_r(c=pos/p_i)=0.5$ vs $P_r(c=pos/p_i)=0.7$ | 0.9985 (0.0000) |

**Table 5.** Performance of IRR ranking with different thresholds values for sentiment classification. (using both IU List-1 and List-2).

| Threshold values for sentiment classification | K=50 | | K=100 | | K=150 | |
|---|---|---|---|---|---|---|
| | Recall (max =0.397) | Precision | Recall (max =0.794) | Precision | Recall | Precision (max=0.840) |
| $P_r(c=pos/p_i)=0.5$ | 0.349 | 0.880 | 0.627 | 0.790 | 0.762 | 0.640 |
| $P_r(c=pos/p_i)=0.3$ | 0.349 | 0.880 | 0.610 | 0.770 | 0.754 | 0.633 |
| $P_r(c=pos/p_i)=0.4$ | 0.349 | 0.880 | 0.603 | 0.760 | 0.746 | 0.637 |
| $P_r(c=pos/p_i)=0.6$ | 0.349 | 0.880 | 0.603 | 0.760 | 0.754 | 0.637 |
| $P_r(c=pos/p_i)=0.7$ | 0.349 | 0.880 | 0.603 | 0.760 | 0.762 | 0.640 |

## 4. Related work

Classic social network theories regarding IUs can be classified into two categories: (1) structure-based centrality metrics, and (2) influence models. Structure-based centrality metrics assess the degree of importance of a node based on the position of the node in a social network. Major centrality measures include betweenness centrality (13). degree centrality (14), closeness centrality (15), and Pagerank in a directed network, in which the *rank* of a node is determined by the rank of those with a link pointing to that node (16). These theories and their extensions have been widely adopted in the analysis of many social networks. The second category of social network theories regarding influential users characterizes the dynamics of social influence using a diffusion model such as the threshold model, the independent cascade model, and their variants and extensions (17, 18). Influential users can then be identified through maximizing the effect of social influence based on these (19).

The emergence of online communities (20), where users often interact through open discussions, provides important opportunities for using novel computational social science approaches (21, 22) to identify influential users from large-scale social networks. In addition to the structure of users' social networks, online communities also capture detailed information of users' online interactions (e.g., the amount, content and time of interactions) that are typically not available in other types of social networks. Research that tries to identify IUs in these communities not only considers network-level centralities, but also incorporates individual users' behaviors and contributions. For example, in online communities that feature contagion or diffusion phenomena, as seen in Twitter or more generally with online viral spreading, researchers used randomized experiments and statistical analysis to find out each individual's influence based on attributes of the individual and the dyadic relationship (23). In an online Q&A community, the difference in knowledge between *question-askers* and *answerers* has been used to find *experts* (24), a type of influential users. Analyses of the blogosphere have utilized blogger contributions (e.g., the number and length of posts) and reader activities (e.g., posting comments to a blog) to assess the influence of a blogger (25).

Meanwhile, social contacts are known to influence health-related behaviors and emotions in individuals (26-28). Provision of emotional support is a key component of OHCs, especially OHCs that cater to individuals with serious medical conditions, such as cancer. Individuals with serious medical conditions often experience stress and anxiety especially around the time of first diagnosis and during treatment (29). However, this important function of OHCs has not been reflected in the literature of IU identification.

## 5. Discussions

This research develops a novel metric that can identify influential users in OHCs. It focuses on the sentimental effect of inter-personal influence on individuals and directly measures an OHC member's ability to influence others' sentiment. This research has important implications for building an active, supportive, and sustainable OHC. For instance, early identification of IUs in an OHC provides community managers an opportunity to publicly recognize their contributions by awarding them prestigious status (e.g., presenting virtual badges of honor) and to encourage other members' participation in the OHC. Also, OHC managers can guide emerging community IUs to assume greater leadership roles and thereby assure consistent, strong peer leadership. This is especially valuable to OHCs, in which influential users are sometimes lost as a result of health-related factors that limit or preclude their continued involvement in the community. The analysis of influential users using IRR illustrated here can also be applied to the study of online social networking sites, such as those focused on product reviews or political opinions, where sentiment is a major part of community interactions.

The proposed metric for influential users is significant not only because it has been shown to be effective for identifying influential users from a large community with a long history, but also because it provides fundamentally new insights into understanding the nature of social influence at multiple scales. The concept of "influential post" introduces, for the first time, a fundamental element of social influence at the inter-personal level. The concept is intrinsically multi-scale, because it is based on the alignment of a responding reply's sentiment (inter-personal level) with the direction of the sentiment change of the thread originator at the individual level. The concept of "influential post" also compliments the previous emphasis on analyzing "relationship" networks with a fundamentally new perspective that analyzes the conversation of actual social interactions in which influence takes place. This new perspective is especially suitable for analyzing influence in online communities, in which interactions emerge and evolve among people previously unconnected. We believe that this new metric will provide an important basis for advancing our understandings about influence, human behaviors, and the dynamics of online communities. For instance, longitudinal studies about influential posts can be useful for studying the dynamic patterns of user engagement in online communities.

As has been the case with other studies of online social influence, our approach is limited in that only inter-personal influence through online interactions is examined. As previously mentioned, the sentiment in a user's post may also be influenced by offline issues. We have tried to eliminate as much offline influence as possible by focusing on the sentiment effect of prompt replies to thread originators. To achieve a complete understanding of influence in an OHC, researchers need to capture and analyze members' offline activities and characteristics. This can then be combined with a more fine-grained text analysis of their posts.

We have mentioned earlier in this paper that another limitation of this work is that it does not distinguish healthy negative sentiment influences (e.g., sadness due to the death of a community member) from those that are not healthy for the community (e.g., opinions that are so strong that can annoy certain community members). Such a distinction, which requires a more fine-grained analysis of the content of the threads, can contribute to the identification of IUs who can have negative impacts on the community. This, we believe, is an important area for future research. Last, in addition to emotional support, we want to measure one's influence in online activities that aim at providing information support. We are also interested gauging one's influence in threads that that do not receive any self-reply.

## 6. Materials and Methods

**6.1 Features and performance for sentiment analysis for posts.**

From posts in the forum, we extract lexical and stylish features for training sentiment models. Table S1 summarizes these features. *Pos* and *Neg* represent the numbers of positive and negative words (and emoticons) respectively. The lists of words' sentiment are based on those introduced by Hu and Liu (30), and the positive and negative emoticon lists are collected from Internet. We found that many posts in CSN forum mention names, e.g., $UserID_x$, *I love the way you think*, as shown in Table 1. Wondering whether name mention has a relationship with sentiment, we introduce NameMention as a feature that counts the occurrences of de-identified names in a post. We also introduce Slang as a feature to check whether the numbers of slang in a post correlate with its sentiment. PosStrength and NegStrength are two features indicate the strength of positive and negative sentiment respectively. Introduced by Thelwall *et al.(31)*, they not only consider whether a word is in positive or negative lists, but also consider the strength (e.g., "*very good*" and *"good!!!"* are more positive than "*good*").

**Table S1.** Features for the sentiment analysis of forum posts.

| Feature | Definition |
|---|---|
| PostLength | The number of words in a post |
| Pos | NumOfPos/PostLength, where NumOfPos is the number of words/emoticons with positive sentiment. |
| Neg | NumOfNeg/PostLength, where NumOfNeg is the number of words/emoticons with positive sentiment. |
| NameMention | NumOfName/PostLength, where NumOfName is the number of names mentioned. |
| Slang | NumOfSlang/PostLength, where NumOfSlang is the number of Internet slangs |
| PosStrength | Positive sentiment strength from the SentiStrength Package. |
| NegStrength | Negative sentiment strength from the SentiStrength Package. |
| PosVsNeg | (NumOfPos+1)/(NumOfNeg+1) |
| PosVsNegStrength | PosStrength/NegStrength |
| Sentence | The number of sentences |
| AvgWordLen | The average length of words |
| QuestionMarks | The number of question marks |
| ExclamationMarks | The number of exclamations |

**Table S2.** Best performance from classifiers.

| Classifier | ROC Area | Classification accuracy |
|---|---|---|
| AdaBoost | 0.832 | 79.2% |
| Logistic regression | 0.832 | 77.5% |
| LogitBoost | 0.816 | 76.8% |
| BayesNet | 0.802 | 74.2% |
| Bagging | 0.794 | 73.5% |
| Neural networks | 0.785 | 73.8% |
| Decision tree | 0.782 | 77.2% |
| SVM | 0.658 | 75.2% |

We try 8 different types of classifiers including AdaBoost, LogitBoost, Bagging, SVM, logistic regression, Neural Networks, BayesNet, and decision trees. All combinations of the features are considered for each type of model to find the features set which achieves the best performance. Both classification accuracy and ROC area are used as metrics because they have different focus on measurement. Classification accuracy measures the percentage of correctly classified instances. ROC area calculates the area under the receiver operating characteristic (ROC) curve, which plots the true positive rate vs. false positive rate for a binary classifier system as the discrimination threshold of the classifier varies. ROC curve can measure the ability of a classifier to pro-duce good relative instance scores, and is insensitive to changes in class distribution. Table S2 summarizes the results of different models based on their best feature sets (10-fold cross-validation). AdaBoost (regression trees are used as weak learners) achieves the best ROC Area (0.832) as well as the best classification accuracy (79.2%). In contrast, the ROC area and accuracy of an AdaBoost using *all* features in Table S1 are even lower--0.813 and 75.2% respectively (too many features may have caused over-fitting).

**Table S3.** Summary of basic features for each community member.

| Group | Features |
|---|---|
| Contribution features | Number of one's initial posts (i.e., posts that start threads) |
| | Number of one's replies to others (i.e., following posts) |
| | Number of threads that one contributed post(s) to |
| | Number of other users' posts published after one's post in the same thread |
| | Avg. response delay between one's post and the next post by others in the same thread (in minutes) |
| | Total length of one's post (in bytes) |
| | Avg. length of one's post (in bytes) |
| | Avg. content length of one's top 30 longest posts (in bytes) |
| | Number of one's active days (one published at least 1 post in an active day) |
| | Time span of one's activity (from first active day to the last) |
| | Avg. number of posts per active day |
| | Avg. number of posts per day during one's time span of activity |
| Network features | One's in-degree and out-degree in the post-reply network |
| | One's betweenness centrality in the post-reply network |
| | One's PageRank in the post-reply network |
| Semantic features | Avg. percentage of words w/ positive sentiment in one's posts |
| | Avg. percentage of words w/ negative sentiment in one's posts |
| | Avg. percentage of Internet slangs/emoticons in one's posts |
| | Avg. percentage of words w/ strong emotion in one's posts |
| | Ratio between the numbers of words w/ positive and negative words in one's posts |
| | Topical diversity (Shannon entropy and log energy of topic distribution in a user's posts) |

**6.2 The classification approach to identify influential users.**

To identify influential users from the OHC, we extract three groups of basic features for each user: contribution, network, and semantic features. Contribution features, as the name implies, measure a user's direct contribution to the forum, such as number of discussion threads (topics) initiated and replies posted, number of days the user has actively posted, length of the user's post, etc. Network features reflect users' centrality (e.g. in/out-degree and betweenness) in a post-reply network, where there is an edge pointing from user A to user B if user A replied a thread started by user B. Semantic features reflect positive or negative sentiment, emotional strength, diversity of topical coverage (utilizing Latent-Dirichlet Allocation), etc. of a user's posts. Table S3 summarizes these basic features. On the basis of these basic features, we also take advantage of the sub-community structure of the social network among community members to generate new neighborhood-based and cluster-based features, which help to improve the performance of our classifiers.

**Table S4.** Performance of various classifiers for identifying IUs.

| Classifier | Top-150 Recall |
|---|---|
| Naïve Bayesian | 0.796 |
| Logistic Regression | 0.706 |
| Random Forest | 0.779 |
| One-class SVM | 0.781 |
| Two-class SVM | 0.739 |
| The ensemble classifier (based on random forest) | 0.854 |

We apply 5 classifiers, Naïve Bayesian, Logistic Regression, and Random Forest, one-class SVM, and two-class SVM, to classify community members into IUs and non-IUs using 10-fold cross-validation. Top-150 recalls (evaluated with IU List-1) obtained from the 5 classifiers range from 0.706 to 0.796 (see Table S4).

Ensemble methods are used to further improve the classification. For each user, a classifier gives a classification result, either as a probability or a binary value to denote whether the user is considered a leader. We then fed each user's five classification results from the five individual classifiers to an ensemble classifier. Among many ensemble methods, the ensemble classifier based on Random Forest achieves the best performance: an average top-150 recall (evaluated with IU List-1) of 0.854 (see Table S4).

**6.3 The list of words and expressions related to death.**

This list was picked from "List of expressions related to death" at Wikipedia[3] and "Death and general words relating to death" at the MacMillan Dictionary Thesaurus[4].

"pass away", "passing away", "passed away", "funeral", "die", "dying", "death", "memorial", "is gone", "was gone", "at rest", "final summons", "room temperature", "at peace", "in peace", "beyond the grave", "beyond the veil", "over the big ridge", "the last roundup", "the great majority", "the ultimate sacrifice", "a last bow", "last breath", "bereavement", "demise", "obituar".

---

[3] http://en.wikipedia.org/wiki/List_of_expressions_related_to_death
[4] http://www.macmillandictionary.com/thesaurus-category/american/Death-and-general-words-relating-to-death